# Detecting Cross-Site Scripting Vulnerabilities through Automated Unit Testing

Mahmoud Mohammadi, Bill Chu, Heather Richter Lipford
UNC Charlotte, Charlotte, NC , USA
(mmoham12, billchu, heather.lipford)@uncc.edu

*Abstract*— **The best practice to prevent Cross Site Scripting (XSS) attacks is to apply encoders to sanitize untrusted data. To balance security and functionality, encoders should be applied to match the web page context, such as HTML body, JavaScript, and style sheets. A common programming error is the use of a wrong encoder to sanitize untrusted data, leaving the application vulnerable. We present a security unit testing approach to detect XSS vulnerabilities caused by improper encoding of untrusted data. Unit tests for the XSS vulnerability are automatically constructed out of each web page and then evaluated by a unit test execution framework. A grammar-based attack generator is used to automatically generate test inputs. We evaluate our approach on a large open source medical records application, demonstrating that we can detect many 0-day XSS vulnerabilities with very low false positives, and that the grammar-based attack generator has better test coverage than industry best practices.**

## I. INTRODUCTION

Cross Site Scripting (XSS) is one of the most common security vulnerabilities in web applications. Cross Site Scripting attacks occur when an attacker successfully injects a malicious JavaScript payload into a web page to be executed by users requesting such a web page. Advised best practice to prevent XSS attacks is to encode untrusted program variables with dynamic content before their values are sent to the browser. While one can prevent all XSS attacks by using the most strict encoder, it also takes away many useful web site functions. To balance security and functionality, developers must therefore choose the appropriate encoder depending on the context of the content, such as HTML or JavaScript. Static analysis [15] techniques are widely used to ensure a web application uses encoding functions to sanitize untrusted data. However, static analysis cannot verify whether the *correct* encoding function is used. Acunetix Web Application Vulnerability Report [1] shows that nearly 38% and 33% of web sites were vulnerable to XSS attacks in 2015 and 2016 respectively.

In this paper we present a unit-testing based approach to automatically detect XSS vulnerabilities due to incorrect encoding function usage. We have built a proof-of-concept implementation for web applications written in Java and JSP. This approach can be extended to other server-side web programming languages as well (e.g. PHP and ASP). For the rest of this section, we provide a brief background on encoding and our approach.

Well tested encoding functions have been written for content placed in the following contexts: HTML body, HTML attribute, CSS, URL, and JavaScript. Consider the fragment of a JSP program shown in Fig 1. Native Java code is enclosed in <% %>.

```
1. <% String pid = (String)request.getParameter("pid");  %>
2. <% String addr = (String) request.getParameter("addr");%>
3. <a href="javascript:void(0)" onclick="action('
   <%=escapeHtml(pid) %> ')"> mylink </a>
4. <p> <%=escapeHtml(addr) %>
```

Fig. 1. Motivating Example

This example has two user provided inputs: *pid* and *addr*. Variable *pid* is used as part of rendering an HTML anchor element on line 3, and *addr* is displayed in the HTML body on line 4. A maliciously supplied input for *addr* might be

<script> atk(); </script>.

If the encoding function, *escapeHtml()*, were not applied, this would cause the execution of the JavaScript function *atk()* on line 4. Encoding function *escapeHtml()* replaces < and > characters with < and > respectively and transforms the malicious input into the following string, preventing *atk();* from being interpreted as a JavaScript program by the browser:

<script> atk(); </script>

However, the same encoding function does not work for the case on line 3. A malicious input for *pid* might be the following

'); atk(); //

It will pass *escapeHtml()* unchanged. The rendered anchor element would be as follows.

<a href= "javascript:void(0)"
onclick= "action( '**');atk(); //** ' ) " > mylink </a>

JavaScript function *atk()* will be executed when the link is clicked. The correct JavaScript encoder would, in this case, replace the single quote character with **\'** to prevent this attack.

There are also cases where more than one encoding function must be used (e.g. an untrusted input used in both JavaScript and URL contexts). The order of applying encoders is sometimes important as well. Unfortunately, there is no systematic way to detect such vulnerabilities.

Other researchers have looked at using type inference to automatically detect the context of an untrusted variable so the correct encoding function can be automatically applied. To aid type inference, such efforts all rely on template languages with stronger type systems, such as Closure Templates [7] or HandleBars [38]. Such approaches have several limitations. First many legacy web applications do not use template languages. Second, type inference is not fully successful even with template languages. For example, a research team from Yahoo! found that they could identify the correct context in about 90.9% of applications written in HandleBars. Other researchers have also shown that type inference is not always accurate for some program constructs written in Closure Templates [32].

Detecting XSS vulnerabilities through black box testing has also been researched, and there are several open source and commercial implementations [5, 9, 22, 30]. In these approaches, a vulnerability is detected by inspecting web application outputs. If an injected attack payload is found in the output, the application is deemed vulnerable. Because the web application output is not rendered by a browser, this evaluation method can have high false positives. Black box testing could also have high false negative rates as well, due to inadequate test path coverage [5].

Our approach is to add testing for XSS vulnerabilities to unit testing, a common software development practice. For example, in iTrust [23], one large open source application we studied, unit test code is available for almost all web pages. Our approach has three components. First, to ensure XSS vulnerability test coverage, we construct multiple unit test files based on one given JSP file in the application. Second, we confirm each vulnerability by rendering attacked pages using JWebUnit. Third, we have a structured way of generating attack strings for test input.

There are several original contributions of this work. We discover vulnerabilities that cannot be found via program analysis. We minimize false positives by confirming vulnerabilities via execution. False positives are a major obstacle for wide adoption of software security tools [5, 30]. Our testing approach can pinpoint exact locations of vulnerabilities, making it easy for remediation. We also minimize false negatives by ensuring path coverage for unit tests as well as systematically generating attack strings using BNF grammars based on modeling how browsers interpret JavaScript programs.

## II. UNIT TEST CONSTRUCTION

To ensure test path coverage, we construct a set of unit tests automatically based on each JSP file with the goal that if the original JSP file has an XSS vulnerability due to incorrect encoder usage, at least one of the constructed unit tests will be similarly vulnerable as well. We refer to the JSP file in the application as the original unit test and each unit test JSP file generated as the XSS unit test. The following are inputs for XSS unit test construction: (1) source code and (2) untrusted sources and (3) sinks. Untrusted sources are Java functions or statements from which malicious data can be brought into the web application, such as request.getParameter(). Sinks are statements used to generate the HTML outputs to be rendered by browsers. There are a number of sinks in the context of JSP applications: *out.write()*, *out.print(), out.println(), out.append()*, or *<%= %>*. We illustrate the unit test generation using Figure 2 as the original code and Figure 3 as one of the constructed XSS unit tests.

To focus our discussions, we assume the application encodes all untrusted variables using known encoding functions. Taint analysis can readily discover cases where an untrusted variable appears in a sink without encoding. Our vulnerability model is a situation where an encoder does not match the application's HTML document context.

A typical JSP file contains both HTML and JavaScript specifications as well as Java variables and statements, referred to as *host variables and statements*. The term *HTML document context* refers to HTML and JavaScript specifications in the JSP file. To avoid false negatives, we capture all sinks in all possible HTML document contexts. For convenience of performing program analysis, we replace all HTML elements with equivalent Java statements. This task is accomplished by using a JSP code analyzer that uses Java output generation commands such as *out.write()* to enclose HTML and JavaScript parts of the JSP files. For example, the HTML elements in line 4 of Figure 2 are replaced by lines 4-6 in Figure 3.

Java branch statements could impact a sink's HTML document context as illustrated by Figure 2. Untrusted variable *ordID* is in a JavaScript context in the "then" branch of a Java if statement (line 4). In the "else" branch of the same Java if statement, variable *ordID* is in an HTML body context (line 6). To ensure we test all possible HTML contexts we test each of the Java if branches in a separate unit test JSP file.

```
1. <% String ordID =    request.getParameter("order");
2. ordID = escapeHtml(ordID);
3. if(editMode){ %>
4.    <a onclick="edit('<%= ordID %>')"  href="#" > Edit Order </a>
5. <% } else { %>
6. <span> Order:<%= ordID %> </span>
7. <% } %>
```
Fig. 2. Original Source Code

```
1.   String ordID = request.getParameter("order");
2.   ordID= escapeHtml(ordID);
//"then" branch of if statemenet
3.   boolean e1= (editMode);
4.   out.write("<a onclick=\"edit(' ");
5.   out.write(ordID);
6.   out.write(" ')\" href=\"#\" > Edit Order</a>");
```
Fig. 3. Generated Unit Test

A control flow analysis is performed to generate the control flow graphs for each JSP file. Multiple XSS unit tests will be created when the JSP file contains *if* or *switch/case* Java statements. For example, the source code of Figure 1 contains the following two possible execution paths:

- Line numbers 1,2,3,4 (then branch)
- Line numbers 1,2,3,6 (else branch)

Two XSS unit tests are generated for this example, each corresponds to one execution path containing no branching logic and each has a sink containing one untrusted variable. Execution paths without sinks and untrusted variables are discarded as they are not vulnerable to XSS attacks.

Figure 3 is a XSS unit test extracted from the "then-branch" of the Java if-statement in Figure 2. The untrusted variable in this unit test is *ordID*, which appears in a sink statement (<%= %>). The sanitizing function is on line 2. To avoid any runtime exceptions or miss any statements affecting the HTML context we also keep the conditional expression used in the if statement in both branches, which is shown as line number 3 of Figure 3, by assigning the value of the conditional expression(editMode) to a Boolean variable *e1*.

While it is possible for branch statements written in JavaScript to change the HTML document context of a sink, we expect such cases to be rare. This is because sinks are part of the application logic written in Java, so it is natural for developers to use Java to express changes in HTML document context. We thus assume that JavaScript code does not change the HTML document context of sinks. We will assess this assumption in the evaluation section.

Figure 4 shows how ZAP [30], a popular security testing tool, conducts tests. The original JSP unit test page is launched on a web server to set up the session. A proxy captures the session information. The captured session is sent as part of subsequent requests by ZAP for security testing. We adopt the same strategy to execute XSS unit tests. Like ZAP, we run the original unit test to set up the session and capture the session id. We then run all the XSS unit tests on the same web server. The captured session id is sent along with attack strings to run the XSS unit tests. Each XSS unit test shares the same session as the original unit test, as session is global across all JSP pages on the server.

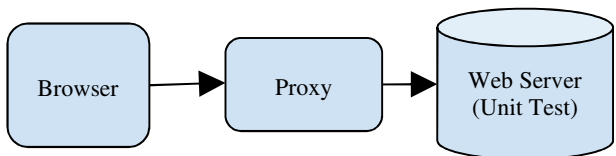

Fig. 4. ZAP's Testing Architecture

**Ideal case.** In the ideal case, the original JSP file contains one untrusted variable as is the case in Figure 2. For such a case, there are no false negatives because all possible HTML document contexts are captured by at least one XSS unit test. If the original code was vulnerable due to using the wrong encoding function, then at least one of the XSS unit tests would be vulnerable.

We define a false positive as a situation where the application's context and the applied encoding function are matched (safe) in the original source code, but the encoding function is detected as vulnerable (mismatched) by a XSS unit test. This is not possible for the ideal case because our XSS unit test construction process preserves the HTML document contexts of the original JSP file.

*A. Injection Points*

We assume that taint sources are specified as a set of Java methods, such as user forms and database queries. Taint flow analysis is used to identify injection points in the program. Injection points are places that the variable containing the attack string (as an input parameter for the unit test) should be used in. This variable is used as the value of the tainted variable that is an argument of the first encoder function. Since a XSS unit test contains no branching logic, detection of such injection points is straightforward. Figure 5 shows part of an original source code. Untrusted variable *fName* is used in a sink on line 5 after being sanitized using two encoders on lines 3 and 4. This variable originated from variable *prf* as result of a database call, *searchProfile()*, a tainted source on line 2. In the corresponding unit test in Figure 6, variable containing the attack string *atk* will be used as the input parameter of the first application of the *escapeHtml()* encoder, on line 3.

We also instrument each XSS unit test so that it reports the line number in the source code if a vulnerability is found as shown in line 6 of Figure 6. We identify the line number of each sink statement in the original JSP file. Suppose the line number of a sink in the original JSP file is L1:

        L1: <%= escapeHTML(x)%>

1. <% List<Profile> prf;
2. prf= searchProfile(customerID);
3. fName = escapeHtml(prf.Name);
4. fName = escapeJavaScript(fName); %>
5. <%= "Full Name:" + fName %>

Fig. 5. Original Code with Injection Point

//atk is a string parameter generated by the test driver
1. atk = request.getParameter("atk");
2. List<Profile> prf;
3. prf= searchProfile(customerID);
// Injection applies to escapeHtml as the value of prf.Name
4. fName = escapeHtml(atk);
5. fName = escapeJavaScript(fName);
// sink line number in original code = 5
6. out.write( addLine("Full Name:" + fName , "5"));

Fig. 6. Generated Unit Test with Injection Point

We add a function to each unit test to add the line number of the sink(L1) to the attack string:

```
out.write(escapeHTML(addLine(x,L1) ))
```

Function *addLine()* adds the line number as a parameter to the attack payload. In our evaluation described below, this line number will be added to the page's title to identify the vulnerable line.

**Multiple Variables:** An XSS unit test may contain multiple untrusted variables. Figure 7 shows two examples. Best secure programing practices [35] suggest that if both variables are properly sanitized with respect to the expected HTML document context, their combination should be safe as well. We refer to this as the *independent encoding assumption*.

This assumption allows us to test one variable at a time by holding the rest of the untrusted variables constant. To avoid runtime errors, we select numeric character 1 as the constant value to be assigned to all other variables while one variable is being tested. This is because 1 can be interpreted either as a character or as a numeric value. We also evaluate the assumption of encoding independence in section 4.

### III. ATTACK EVALUATION

The goal of attack evaluation is to assess whether an XSS unit test is vulnerable to any of the XSS attack strings. One widely used security testing approach, exemplified by the popular black box testing tool ZAP [30], is to look up the attack payload in the response page. The rationale for this approach is that if an attack payload can bypass the encoder function intact, an attack could occur. Unfortunately this approach can lead to high false positives. A successful attack payload must be compatible with the context it is injected into.

For instance, Figure 8 shows a situation in which an HTML body encoder is used to sanitize a user-entered parameter on line 1. Line 2 is an attack string from ZAP's attack repository. Line 3 shows a part of the output of the web page when this attack string is applied. Since the encoder does not alter the attack string, ZAP's test evaluation mechanism would report this page as vulnerable. This is a false positive because this attack cannot be executed in the HTML context.

1) <%= escapeHtml(var1) + "constant" + escapeHtml(var2) %>

2) <%= escapeHtml(var1+ "constant" + var2) %>

Fig. 7. Multiple Variables

1. <div> <%= escapeHtml(request.getParameter("atk"))%> </div>

2. Attack String : + alert(1)
3. <div> + alert(1) </div>

Fig. 8. False Positive in Attack Detection

### A. Attack payload

Our approach is to execute unit tests in headless browser components such as JWebUnit, a widely used testing tool for web pages. Vulnerabilities are only reported if successful execution of an attack payload by JWebUnit is detected. For the attack payload, we use a JavaScript function *attack(n),* which takes as parameter the line number of the sink statement being tested. It changes the web page title by appending that line number.

### B. Test driver

Figure 9 shows the XSS unit test driver. Lines 1 and 2 are for test preparation. Function *sessionPreparation()* sets up the execution environment by applying captured session information. The rest of the test driver invokes the XSS unit test by applying attack strings. After initializing an instance of WebTester (a subclass of JWebUnit) on line 1, each iteration of the loop on line 2 takes one attack vector (*atk*) and invokes the XSS unit test page (*unitTest_1.jsp*) with the attack string as a parameter (line 4). Line 5 pauses to let the unit test page be rendered completely. Line 6 asserts whether the attack is successful by checking the title of the response page. If the attack is successful, the page title contains line number(s) of the vulnerable sinks, helping developers to fix vulnerabilities.

### C. Handling events

In order to find vulnerabilities associated with tag events, we must trigger each event with a JavaScript body. There are near 88 possible events in HTML5, some of them can only be triggered based on particular user interaction such as *onmouseover* or a run time condition such as *onerror*.

```
Public void prepare()
{
1.   wt = new WebTester();
2.   sessionPreparation(); //other preparations such as proxy
}
public void run() {
3.   for( String atk :atkVectors){
// Invoking the Unit Test
4.     wt.gotoPage("unit_1.jsp?atk="+atk);
5.     sleep(100);
6.     verifyResponse(wt);
  }
}
```

Fig. 9. Test Driver with Test Preparation

```
1. var tags = document.all;
2. for (var i=0; i <tags.length;i++){ e= tags[i];
3.   if (typeof e.onfocus == "function")  {
4.     event = e.onfocus;
5.     e.onclick=event;
6.     e.click();   }
 … check for other events
```

Fig. 10. Triggering all events having a handler

However, since all events share the same syntax, we can substitute events that cannot be easily simulated in a test environment with an event that can be easily triggered.

We verified that in major browsers (Chrome, Safari, Firefox) event *onclick* can be associated with every HTML tag and it can be triggered using a JavaScript API. Figure 10 shows a JavaScript program we use to go through all tags in the DOM. For each tag, the program checks if the tag has an event with an event body (line 3). If a tag has a body, the program assigns the event body to an *onclick* event and triggers it programmatically.

## IV. ATTACK GENERATION

Because our test evaluation is based on execution of attack strings, we must make sure attack strings are syntactically correct. Furthermore, we want to include all possible types of attack scenarios. Related work in generating XSS attacks relies on either expert input [21], or on reported attacks [22, 37]. It is difficult to show that all possible attack scenarios are included using these approaches.

Our approach consists of two components. First we use grammars to model how JavaScript payloads are interpreted by a typical browser. Assuming this grammar is accurate, then a successful attack must follow the rules of this grammar. Second, we devise an algorithm to derive attack strings systematically based on the grammar. Assuming the grammar accurately models the way the browser interprets JavaScript, and assuming that the attack derivation algorithm can generate at least one attack string for every type of attack, then our approach would cover all possible attack scenarios. It is possible that either the grammar may have missed a way by which a browser interprets JavaScript, or the attack enumeration algorithm failed to consider a possible derivation path based on the grammar. Through peer review, we can improve both components in a way similar to how crypto algorithms are revised.

### A. Attack Grammars

A typical web browser contains multiple interpreters: HTML, CSS, URI and JavaScript. The browser behavior can be modeled as one interpreter passing control to another upon parsing specific input tokens while rendering HTML documents. We refer to the event of interpreter switching as *context switching*. For example, the URI parser transfers the control to the JavaScript parser if it detects input *javascript:* (if supported by the browser) as in the case:

A successful XSS attack is to induce the JavaScript interpreter to execute the attack payload. We use a set of context free grammar (CFG) rules to specify possible input strings that cause the browser to activate the JavaScript interpreter to execute an attack payload. Portners et. al. [31] observed that a successful XSS attack must either call a JavaScript function (e.g. an API), or make an assignment (e.g. change the DOM). In JavaScript, wherever an assignment operation can be executed, a function call can also be made. Therefore, without loss of generality, we assume the attack payload is a function call *atk()* that changes the title of the webpage.

Like Halfond et. al [19], we divide the CFG into different sections: URI, CSS, HTML, Event and JavaScript. In each section we specify possible transitions to cause a JavaScript interpreter to execute an attack payload. We will then integrate these contexts to generate attack strings. Throughout the paper we will use the following convention: upper case words for non-terminals, lower case words for terminals, symbols *sq*, *dq*, *rp*, *lp*, *eq* for single quote('), double quote("), right parenthesis, left parenthesis and equal sign characters respectively.

*1) URI context:*
URI (Uniform Resource Identifier) strings identify locations of resources such images or script files. Based on RFC 3986, they have the following generic syntax:

scheme:[//[user:password@]host[:port]][/]path[?query][#fragment]

Here, the scheme represents the type of protocol (such as ftp or http) used to access a resource, and the rest of the string expresses the authority and path information required to identify the resource. To cause the URI interpreter to switch to the JavaScript interpreter, the scheme must be equal to the keyword *javascript*, followed by JavaScript statements. Other possible schemes include http, ftp, and https. Since no JavaScript can be injected into schemes other than scheme *javascript*, we concentrate on describing URIs that contains the JavaScript scheme [11]. An URI can be properly interpreted by a browser only as a value of an expected attribute of a host context. We continue with the example of

,

where *src* is the source attribute of the HTML *img* tag. It is referred to as URIHOST. Figure 11 represents the grammar for URI. Rule URIATRIB specifies a URI attribute consisting of a URIHOST name and the URLVAL. Rule URIHOST lists all possible URI host contexts in an HTML document. Again, for the purpose of generating attack strings, we only consider a URI of the JavaScript scheme. PAYLOAD is a special nonterminal representing a JavaScript attack payload. It signals to our attack generator that a context switch to JavaScript is possible at this point.

```
<URIATRIB> ::= URIHOST  eq  URIVAL
<URIHOST>  ::=  src  |  href  |  codebase  |  cite|action  |
background  |  data  |  classid  |  longdesc|profile  |usemap  |
formaction|icon | manifest | poster | srcset | archive
<URIVAL>::= sq URI sq | dq URI dq | URI
<URI> ::= javascript: PAYLOAD
```

Fig. 11. URI Grammar

*2) CSS Context:*
Cascading Style Sheets (CSS) specifications can be either contained in a CSS file or placed directly in HTML elements,

e.g. tag definitions (using the style attribute or <style> blocks). A context switch from the CSS interpreter to the JavaScript interpreter is possible only when a URI is a property of a CSS-style element, specified by function url(). Figure 12 lists rules for URI to be included as part of a CSS-style element, starting with tokens style=. Rule STYLE defines a CSS style as a list of properties composed of name PROPNAME and value PROPVAL. Rule PROPNAME lists all possible properties requiring URI strings. PROPVAL specifies how function url() uses URI non-terminals to include URI context.

*3) Event Context:*

HTML events, such as onfocus and onload, can cause context switches to JavaScript. Grammar rules in Figure 13 define an HTML event attribute composed of an event name EVENTNAME and value EVENTVAL. Although types of possible events vary with HTML tags, we found that the onclick event can be triggered in all HTML tags. As mentioned previously, we change all events in the source code to the onclick event. Rule EVENTVAL defines the value of the event which is a JavaScript statement to be executed upon the specified event, as represented by the special nonterminal PAYLOAD.

*4) Integrating the contexts into HTML.*

Having modeled context switches in URI, CSS, and Event, we integrate them in an attack grammar for HTML in Figure 14. Rule HTML defines tags as a set of elements represented by the ELEM rule. Some tags cannot have tag bodies (such as *img*, *input* or *param* tags). Tag script and style are special in that their entire body is interpreted as JavaScript and CSS respectively. Tags *textarea* and *title* can only contain text characters in their bodies, thus JavaScript programs cannot be injected. These special tags are reflected in ELEM rules defining these tag categories using corresponding rule names. We use rule IMG to define tag *img* as a representative to model the rest of tags, as they all share the same syntax. This rule defines an *img* tag as a tag name followed by a list of attributes ATRIBULIST. In ATTRBLIST, the browser can switch to the JavaScript interpreter depending on the attribute type: one of URI, CSS, and EVENT. Attack grammars for these cases have been described in detail earlier.

*5) JavaScript.*

JavaScript code can be placed either directly in HTML elements (e.g. through tag events such as onclick) or in <script> blocks. Attackers can inject a malicious payload into a block of vulnerable JavaScript code. When attacking the JavaScript context, no interpreter switch is needed, a successful attack must manipulate the JavaScript interpreter into executing the payload, attack().

Injection points in JavaScript are (Java) host variables. While host variables could be used in any JavaScript construct, such as part of a variable or function name (e.g., var *idx<%= varName%>* = *'value';*), such cases make little sense. Host variables are primarily used to pass server-side values to JavaScript code. Thus we only consider scenarios where attack scripts are injected as part of a string or a numeric literal in expressions as illustrated by the following examples.

```
<STYLEATRIB> ::=   style eq STYLEVAL
<STYLEVAL> ::= (sq STYLE sq) | (dq STYLE dq') | (STYLE)
<STYLE> ::= CSSPROP*
<CSSPROP> ::= PROPNAME : PROPVAL;
<PROPNAME> ::= background-image | list- style-image| content | cursor | cue-after | cue-before
<PROPVAL> ::= url(URI)
```

Fig. 12. CSS Grammar

```
<EVENTATRIB>::= EVENTNAME eq EVENTVAL
<EVENTNAME>::=onclick
<EVENTVAL> ::= sq PAYLOAD sq | dq PAYLOAD dq | PAYLOAD
```

Fig. 13. Event Grammar

```
<HTML> := ELEM*
<ELEM> := TEXT | STYLE | SCRIPT |IMG
<TEXT> := TITLE | TEXTAREA
<TAREA> := <textarea> CDATA </textarea>
<STYLE> := <style> CSSPROP* </style>
<SCRIPT> := <script> PAYLOAD </script>
:= 
<ATRIBLIST>::= ATTRIBUTE*
<ATTRIBUTE> ::= URIATRIB | STYLEATRIB | EVENTATRIB
```

Fig. 14. Integration Grammar.

| | |
|---|---|
| "20<%= **hostVar** %>" | : Part of String |
| 20<%= **hostVar** %> | : Part of Number |
| 20 * <%= **hostVar** %> | : Operands |
| func( "a",<%= **hostVar** %>, 1) | : Parameter |

The goal of the attack is to turn the host variable into an expression so a function call can be made. A successful attack can be any syntactically correct expression. Without loss of generality, we generate attack expressions using only the + operator. The resulting expression is referred to as an additive expression. Its grammar is shown in Figure 15. The first two lines in Figure 15 define additive expressions as expressions composed of multiple string/numeric literals or expressions concatenated to each other using the plus(+) operator. PAYLOAD non-terminal is a placeholder for attack payloads.

```
AdditiveExp = PrimaryExp AdditivePart
AdditivePart = (+ PrimaryExpr)*
PrimaryExp = PAYLOAD | Literal
Literal = dq 1 dq | sq 1 sq | 1
```

Fig. 15. JavaScript Additive Expression Grammar.

## B. Attack String Generation

The goal of the attack string generation is to generate all possible types of attacks using the attack grammar. Space limitations only permit us to describe our algorithm through a series of examples in this section.

### 1) Sentence Derivation

We generate XSS attack strings based on any of the attack grammars described above by constructing a leftmost derivation tree [20] from the start symbol of each grammar. The following code shows the steps of deriving an attack string based on the HTML image tag.

ELEM = IMG
    ::= <**img** ATTRIBUTE*>
    ::= <**img** EVENTATRIB >
    ::= <**img** EVENTNAME eq EVENTVAL>
    ::= <**img** onclick = **"PAYLOAD"** >

### 2) Generating attack strings

Attacks can be injected in any part of an HTML element, a CSS block, or JavaScript expression. Consider the following example where a host variable, *InjPoint,* is in the value part of the *fName* variable as part of a function parameter.

    **var fName =func("Dr.** <%= InjPoint %> **");**

The attack script must take into consideration the existing context both to the left and to the right of the injection point. These existing contexts are referred to as *left context* and *right context* respectively as described in our previous work [28]. They are shown in **bold** above. To fit the attack into the left context, one may close the string parameter with character **"** followed by a context switch using a new expression. The resulting attack string would be:

    *" + attack() +"*

Furthermore, the attack must consider existing context to the right of the injection point by relying on existing right context to complete the attribute element. The successful injection is shown as follows.

    var fName =func("Dr**." +attack()+"** ");

If we regard derivations starting from the start symbol of the grammar as complete sentences of the grammar, to systematically generate attacks for all possible existing left and right contexts, we must produce all possible partial sentences. The following is a possible derivation for an additive expression in JavaScript leading to a complete sentence:

    AdditiveExp = PrimaryExp AdditivePart
    … ::=Literal AdditivePart
    … ::= "1" AdditivePart
    … ::= "1" (+ PrimaryExpr)*
    … ::= "1" + PrimaryExp + PrimaryExp
    … ::= "1" + PAYLOAD + "1"

For each complete sentence derived from the attack grammar, we generate multiple versions of partial sentence as potential attack strings.

Each version will be shaped by removing one token from the either the beginning or from the end of the previous version starting from the initial sentence. These versions represent different possible ways an attack can be successfully interpreted by the browser taking advantage of the injection point's left and right contexts.

To consider existing contexts to the left of the injection point we generated different partial sentences by removing one token from the beginning of the previous version. This removing process will continue until the first *payload* symbol is reached. For example, given the additive expression derived earlier, the following versions of attack strings can be generated.

    Initial Sentence: "1" + PAYLOAD + "1"
    1" + PAYLOAD + "1"
    " + PAYLOAD + "1"
    + PAYLOAD + "1"
    PAYLOAD + "1"

To consider existing contexts to the right of the injection point, we systematically generate multiple versions of any attack string by removing one token from the end of the previous one until the *payload* symbol is reached. Then the following four more versions of attack strings will be generated based on the attack string (" + PAYLOAD + "1"):

    "+PAYLOAD + "1

    **"+PAYLOAD + "**

    "+PAYLOAD +

    "+PAYLOAD

The second version can be successfully injected into host variable *InjPoint*.

### 3) Closures

Closure operators (*, +) in our attack grammar may result in an infinite number of derivations. The following example shows a simple derivation by applying the closure operator up to two times( * = 2). A total of six derivations are possible.

HTML := ELEM*
ELEM := (IMG | SCRIPT)*
    := IMG
    := IMG IMG
    := IMG SCRIPT
    := SCRIPT
    := SCRIPT SCRIPT
    := SCRIPT IMG

To generate a finite set of attack vectors, we observe that attack scripts containing more than one attack payload are redundant. This is because a successful attack only needs to execute one payload. For a set of attack grammars, we empirically determine how many times closure operators need to be applied. We compute leftmost derivations by applying different upper bounds on closure operators until no more attack strings with one payload can be added. For example, for the grammars presented here, applying each closure operator 3 times does not generate new attack strings with one payload over applying each closure operator 2 times. Note that current attack grammars do not contain recursive rules. We are examining how to handle recursive rules.

In summary we generate an initial set of attack strings by deriving sentences from the start symbol of each grammar (URI, CSS, HTML, EVENT, JavaScript). Each closure operation is applied up to two times. For each attack string in the initial set, we generate additional versions of the attack string by dropping tokens from both the left and right as described above. Only attack strings with a single context-switch are included in the final set of attack strings for unit testing. Because a successful attack script only needs one working context-switching mechanism and all possible combinations of context-switching are generated, thus combining them in one attack script is not required.

## V. EVALUATIONS

Our evaluation uses iTrust, an open source medical records application with 112,000 lines of Java/JSP code [23]. We seek to evaluate (1) the performance of our XSS vulnerability detection, (2) the completeness of our attack generation approach, and (3) the computational performance of our approach. Project iTrust has 235 JSP files and we randomly selected ⅓ or 70 JSP files for this evaluation.

### A. Vulnerability Detection

We compared our XSS unit testing approach with security black box testing using the popular open source security testing tool ZAP. Table 1 summarizes our evaluation results.

Table 1: Summary of vulnerability findings

|  | Vulnerabilities reported | True Positive | False Positive |
|---|---|---|---|
| ZAP | 31 | 9 | 22 |
| XSS Unit Testing | 17 | 17 | 0 |

We found 17 zero-day vulnerabilities in 13% of JSP pages tested. Most vulnerabilities are due to the use of HTML encoding in JavaScript context, as shown below:

`<a onclick="func('<%= escapeHtml(input) %>')" ...>Link</a>`

ZAP has a very high false positive rate: 70%. Our approach has no false positives. The reason for ZAP's high false positive rates is because it does not confirm test result via execution, as illustrated in Section III Figure 8. False positives for our case are still possible as illustrated by the code in Figure 16 where a JavaScript encoder should have been used. However, in this case statement containing substring() has the side effect of sanitizing the output making this situation not exploitable.

```
zip = request.getParameter("zip")
zip =zip.substring(0, 5);
<a onclick=" func(' escapeHTML(<%= zip% >)' ); " ... >
```

Fig. 16. Possible False Positive.

Our approach found 8 vulnerabilities ZAP did not find. All these cases are due to lack of test coverage by ZAP. ZAP did not test all execution paths. In our approach, a separate XSS unit test for each possible execution branch in the original JSP file will be evaluated.

In addition, we investigated iTrust for any JavaScript flow control statements causing different HTML contexts at client side. We could not find any JavaScript code inducing vulnerabilities that cannot be detected. We examined iTrust for sinks having multiple variables. Out of 2268 sinks, 27 contain multi-variables. In all sinks with multiple variables, our encoding independence assumption is true. That is to say, encoding each variable separately adequately allows us to test for XSS attacks.

### B. Attack generation

We compare our grammar based attack generation with two well regarded open source XSS attack repositories: ZAP and HTML5 Security web site [21]. We applied these attack repositories to the XSS unit tests we constructed. Both ZAP and HTML5Sec attack repositories found the same vulnerabilities. However, we found several vulnerabilities that cannot be detected by ZAP and HTML5 Security cheat sheet attack repositories. One example is shown below.

`<div style="height: <%= escapeHtml(input) %>px; "> </div>`

The following attack string generated by our approach can detect this vulnerability.

`;background-image:url('javascript:attack()');`

Attack repositories in ZAP and HTML5Sec rely on contributions from pentesting experts. Our approach systematically derives attack strings based on a set of attack grammar.

### C. Computational performance

We looked at the performance of XSS unit testing using experiments performed on a desktop Mac with a 2.7 GHz Intel core i5 with 8GB RAM. Our attack generator produced 223 attack strings, which were applied to each unit test. For the iTrust pages we sampled, the average time to evaluate each XSS unit test against all attack vectors was 17 seconds.

JSP files may contain multiple execution paths but only those containing sinks with tainted variables will be tested. Our evaluation of 70 JSP pages in iTrust shows that on average a JSP file leads to 29 XSS unit tests. Average execution time to complete XSS vulnerability testing for a JSP file is thus 493 seconds or 8.2 min. Generation of XSS unit tests is much faster than running all the tests. This performance is reasonable for our testing environment and can scale to a real world environment.

## VI. RELATED WORKS

Researchers have investigated a variety of prevention and detection techniques to mitigate XSS vulnerabilities. Preventive approaches include secure programming guides to inform the developers to use encoding functions correctly. Well known guidelines include the OWASP XSS cheat sheet [36] and best practices by Graff and Wyk [16].

Attempts have been made to automatically sanitize untrusted input using template languages. We discussed approaches for auto sanitization via type inference in section I [32, 33]. Technological restrictions, such as the use of template languages, means such approaches are not widely applicable to many legacy web applications.

Johns et al. [24] have developed an abstract data type that strictly enforces data and code separation in a host language such as Java. However, this approach comes with a significant, 25%, run-time overhead. ScriptGard [33] is a run-time auto-sanitization technique in ASP.Net similar to [18] for Java and WASP for SQL injection[19]. Advantages of these approaches are that they can automatically sanitize large scale legacy systems using a path-sensitive approach using binary code instrumentation. All these approaches require a runtime component that could incur runtime overheads. Furthermore, requiring a runtime component necessitates changes to existing infrastructures, such as browsers. Our approach works with all web languages (HTML, CSS, JavaScript) and requires no runtime support.

Static analysis techniques are widely used to detect XSS vulnerabilities using taint analysis techniques [3, 17, 34, 39]. The main disadvantage of static analysis is high false positive rates [15]. Furthermore, static analysis tools can only check for the existence of the sanitization functions and not evaluate their effectiveness [4]. Thus, static analysis cannot be used to address the context-sensitiveness of sanitization errors [2].

Dynamic analysis techniques aim to evaluate applications' responses to detect any sanitization mistakes [12, 14, 26]. In the case of the black box testing, different algorithms such as combinatorial testing [6, 13], pattern-based algorithms [8, 25], and attack repositories [34] have been explored. Duchene et al. in [8, 10] proposed a control- and data- flow aware fuzzing technique. They use a state-aware crawler to record application requests and responses and use them to infer an application's control and data flow. Fuzzing process is guided by this information. The advantage of data flow inference is that it enables more accurate detection of stored XSS vulnerabilities. Because crawling-based inference is source-code independent and it uses automatic form filling and pruning techniques, the inferred control flow may not be complete, leading to potentially high false negatives. In contrast, we utilize source code analysis to extract all execution paths to avoid missing any sinks.

McAllister et. al [27] proposed an interactive black-box vulnerability scanner. They aim to increase test coverage by leveraging user activities through guided fuzzing. However, relying on the user's activities to increases test coverage is not complete and can lead to false negatives.

There are other types of XSS sanitization functions (other than encoders) not addressed by our approach. Consider a blogging web site allowing use of HTML markup tags as input. HTML encoding functions are not proper here because they would disable all HTML markup tags. There are heuristic filters, e.g. [29], that try to block unwanted JavaScript programs in HTML body context. Such filters are difficult to verify as many patches have been issued.

## VII. CONCLUSION

In summary, we propose a unit testing approach to detect cross-site scripting vulnerabilities caused by incorrect encoder usage. This approach can be easily integrated into existing software development practices and can pinpoint the location of a vulnerability in the source code. It can help developers find and fix XSS vulnerabilities early in the development cycle, when they unit test their code, without involving security experts. The grammar-based attack generation is a structured way to generate XSS attack strings. We were able to generate tests for vulnerabilities missed by popular attack repositories. More importantly, our grammar models can be modified to cover unknown or new attack scenarios. For example, a new version of a browser may offer new ways for attackers. Our approach also has low false positive rates. Finally, our evaluation demonstrates that our approach is computationally efficient and can detect vulnerabilities cannot be found using black-box fuzzing systems.

This work can be extended in a number of ways. We are currently evaluating our approach with more open source applications. We also plan to look at handling recursive structures in the attack grammar as well as improve the efficiency of attack evaluation.

## VIII. ACKNOWLEDGMENT

This work is supported in part by the following grants from the National Science Foundation: 1129190, 1318854.